\documentclass[
 reprint,prb,
 amsmath,amssymb,
 aps,
]{revtex4-2}
\usepackage{graphicx}
\usepackage{bm}
\usepackage{braket}

\begin{document}
\title{Missing link between the 2D Quantum Hall problem and 1D quasicrystals}

\author{Anuradha Jagannathan}
\affiliation{
Laboratoire de Physique des Solides, Universit\'{e} Paris-Saclay, 91405 Orsay, France
}
\date{February 2023}

\begin{abstract}
This paper discusses a connection between two important classes of materials, namely quasicrystals and topological insulators as exemplified by the Quantum Hall problem. It has been remarked that the quasicrystal ``inherits" topological properties from the 2D Quantum Hall model. The aim of this work is to show this explicitly by introducing the Fibonacci-Hall model as a link between a 1D quasicrystal and the magnetic problems. We show here how Chern numbers for bands in periodic approximants of quasicrystals can be computed, along with gap labels. The Chern numbers are thus seen as a consequence of a flux parameter $\phi^S$ induced by the geometry of winding in 2D space of the quasicrystal. We show the existence of lines of Lifshitz transitions in the phase space of the model. These are marked by change of Chern number and in open systems by disappearance of edge states. Our extrapolation method can be generalized to higher dimensional 2D and 3D quasicrystals, where higher order Chern numbers could be computed, and importantly, related to experimentally measurable transport quantities.  
\end{abstract}

\maketitle
\section{Introduction}
It is well-known that, while quasicrystals have no periodicity in real space, they can be obtained by projecting down from a higher dimensional periodic lattice. Topological band structures in quasicrystals have been a topic of active study in recent years. Most of these studies discuss topological properties which are $extrinsic$ -- for example, induced by magnetic field, or via spin-orbit and other couplings \cite{fuchs,duncan,huangliu,chen,cao,ghadimi,kobialka}. In contrast, this paper considers the $intrinsic$ topological properties of quasicrystals, arising solely due to their structure. A connection between the simplest quasicrystal, the Fibonacci chain and the 2D Quantum Hall (QH) problem embodied by the Harper \cite{harper} and Aubry-Andre models \cite{aamodel} has been pointed out \cite{kraus} and generalized \cite{ganeshan}. While such a relation exists formally, a correspondence between the 1D and 2D problems has not so far been made explicitly. This paper aims to fill the gap by introducing the Fibonacci-Hall model, as a candidate for the missing link between the two problems. It provides a new way to compute and understand topological properties governing physical properties of quasicrystals.

The class of quasiperiodic Hamiltonians covered in this paper are tight-binding single electron models with diagonal and off-diagonal terms of the form
\begin{eqnarray}
\label{eq:ham1D}
    H^{1D} &=& - \sum_{\langle i,j\rangle} t_{i} (c^\dag_i c_{i+1} + h.c.) + 
 \sum_i \epsilon_i c^\dag_i c_i 
\end{eqnarray}
where the parameters $t_{i}$ and $\epsilon_{i}$ vary in space quasiperiodically. The spin indices have been suppressed as they do not play a significant role in our discussion. 
For almost periodic 1D systems, which include the above models, there is a ``gap labeling theorem" due to Bellissard and coworkers \cite{bellissardgap} which states that all their spectral gaps can be indexed by integers -- the gap labels. In the last decade or so of research on topological quasicrystals, the meaning of these gap labels has been clarified, and they have even been observed experimentally \cite{tanese,dareau}. In particular, the topological equivalence between the Fibonacci and the Aubry-André-Harper models was shown, and used in a topological pumping experiment \cite{kraus}. We show by explicit construction of the Fibonacci-Hall Hamiltonian, that bands and gaps evolve continuously from the 2D QH band structure to those of the 1D quasicrystal.  For finite periodic approximants, band Chern numbers are exactly found by a standard method \cite{loring}. We show that the gap labels deduced from the band Chern numbers are consistent with known values deduced using the gap labeling theorem. The phase space of the model has Lifshitz type transitions. In the open system, we show that edge states appear in the topological region of the phase diagram and that they disappear at the topological transition into the trivial state.  

The quasicrystal-Quantum Hall correspondence can be readily generalized to other 1D quasicrystals. It could be extended to higher dimensions, allowing for direct computations of higher order Chern numbers, with implications for experimentally measurable transport coefficients.

\section{The Fibonacci, Quantum Hall and Fibonacci-Hall models}
\begin{figure*}
\includegraphics[width=0.2\textwidth]{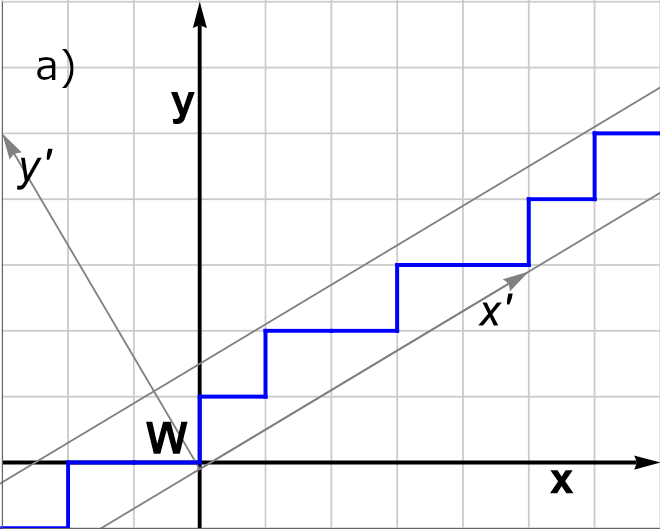} \hskip 0.5cm
\includegraphics[width=0.2\textwidth]{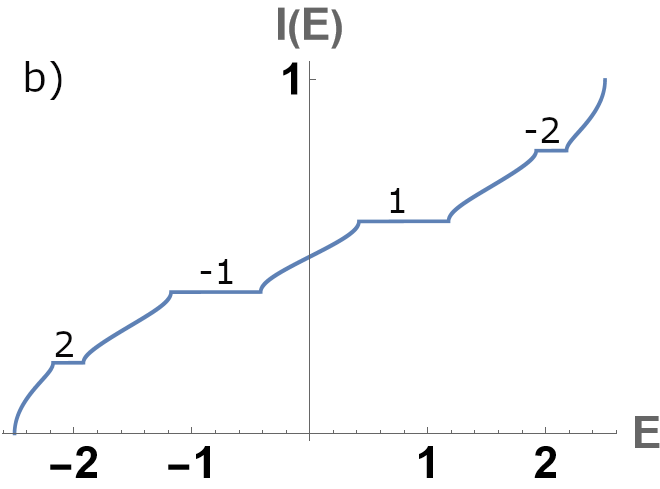} \hskip 0.5cm
\includegraphics[width=0.23\textwidth]{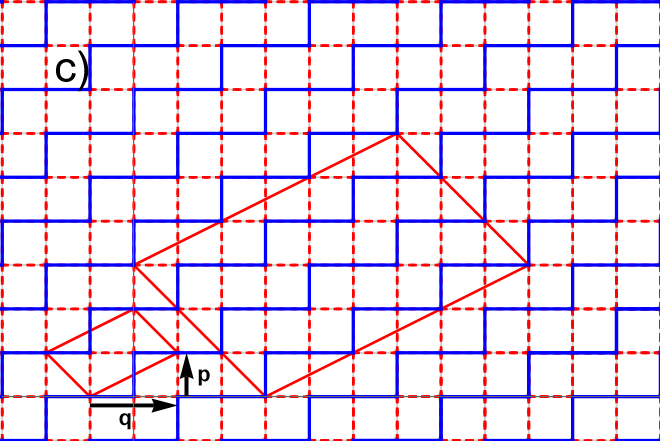} \hskip 0.5cm 
\includegraphics[width=0.15\textwidth]{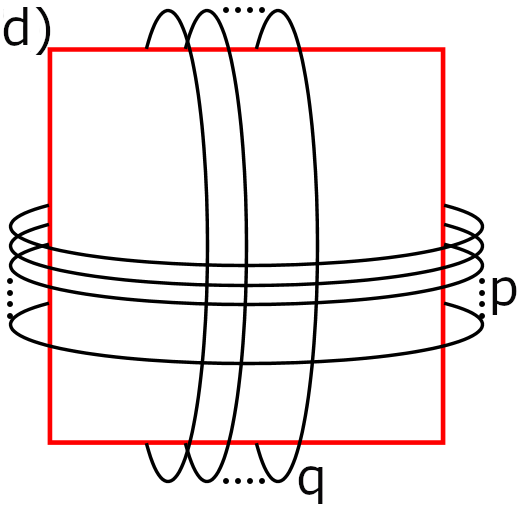}
\caption{a) Schema of the cut-and-project method. Selected points (joined by the zigzag blue line) lying in a strip $\mathcal{S}$ of a 2D square lattice. These points are projected onto the $x'$ axis to form the Fibonacci quasicrystal. The interval $W$ represents the perpendicular cross-section of the strip corresponding to one unit square. b) Integrated DOS for the $k=4$ chain, showing the four bands, along with their gap labels deduced from Eq.3.  c) The figure shows the parallel stacking of approximants in 2D for the approximant $N=3$ ($p=1,q=2$). A structural unit cell and the bigger magnetic unit cell are shown outlined in red. Intrachain bonds are colored blue, interchain bonds are colored red. d) Representation of an approximant chain, showing the wrapping of the rational cut around the unit torus, with winding numbers $q$ and $p$ along the $x$ and the $y$ directions respectively.  }
\label{fig:cutproj}
\end{figure*}

In the cut-and-project method, the Fibonacci chain is obtained by projecting a set of points of the square lattice. These are points that lie within a strip $S$ of slope of $1/\tau$, where $\tau=(\sqrt{5}+1)/2$ is the golden mean, and perpendicular cross-section $W$, as shown in Fig.\ref{fig:cutproj}a). The quasicrystal is had on projecting the selected points onto the $x'$ axis (the direction of the strip). The slope may instead be a rational given by $F_{k-2}/F_{k-1}$, where $F_k$ is a Fibonacci number. This generates a periodic approximant having a unit cell of $F_k$ sites, and the quasicrystal is obtained in the limit $k\rightarrow \infty$.  The figure is drawn for a strip which passes infinitesimally below the origin $O$, but its position is arbitrary. If the strip is displaced parallel to itself, the zigzag blue line of selected points changes (``phason flips" occur). The origin of the unit cell is thereby changed, but the overall structure is not. Taking the vertical (resp. horizontal) hopping amplitudes to be $t_b$(resp. $t_a$) leads to a pure hopping model of the family $H^{1D}$, namely

\begin{eqnarray}
\label{eq:hamFibo}
    H^{F} &=& - \sum_{i} t_{i} (c^\dag_i c_{i+1} + h.c.) 
\end{eqnarray}
where $t_i$ is equal to $t_a$($t_b$) for horizontal (vertical) bonds and the unit cell consists of $F_k$ sites. 
In the following we will focus on this model, referred to as the ``off-diagonal" Fibonacci model, but the approach could be modified to obtain other forms of $H^{1D}$. Many important results have been obtained for the Fibonacci models using their invariance under scale transformations combined with renormalization group (reviewed in \cite{jagaRMP}). The spectrum of the model Eq.\ref{eq:hamFibo} has a hierarchical structure. The energy levels of the $k$th approximant have subgroups of levels, which in turn are composed of subgroups and so on. Spectra of larger approximants can be related to those of smaller approximants. The resulting gap structure is also hierarchical. 
The gap labeling theorem, as adapted for the $k$th periodic approximant \cite{macegaplabels}) states that the integrated density of states within the $j$th gap, $I^{(k)}_j$, can be written as
\begin{eqnarray}
\label{eq:gaplabeling}
I^{(k)}_j = \left[g_j\frac{ F_{k-1}}{F_k} \right], 
\end{eqnarray} 
where $g_j$ is an integer ($\vert g_j \vert \leq F_k/2$) and where $[.]$ denotes the fractional part. Given any approximant all its gap labels can be readily computed using this formula. Fig. 1b) shows the IDOS for the $k=4$ (5 site) chain with $t_a=1,t_b=1.6$, showing its gaps and their labels $g_j$ ($j=1,..,4$) obtained using Eq.3.

Now, by extending the cut-and-project procedure to an infinite set of parallel strips, the entire 2D lattice can be regarded as an infinite set of parallel chains. Fig.\ref{fig:cutproj}c) shows two unit cells of rhombic shape -- the smaller rhombus is a structural unit cell for the $k=3$ approximant, while the larger rhombus is that of a magnetic cell as discussed below. Blue bonds connect sites along the $same$ chain, and red bonds connect sites on neighboring chains. 

A 2D Quantum Hall Hamiltonian $H^{2D}$ is now defined. To reproduce the Fibonacci winding property, a fictitious dimensionless flux $\phi^{S} =Ba^2/\phi_0$ is introduced, where $B$ is a fictitious magnetic field perpendicular to the $x-y$ plane, $a$ the side length and $\phi_0$ the flux quantum. The superscript ``S" indicates that this flux arises due to the winding property induced by the strip $S$. The flux $\phi_S$ has a different rational value for each approximant, and is set by the winding of the strip in the 2D lattice. Fig.\ref{fig:cutproj}d) is an equivalent compact representation of an periodic approximant corresponding to a slope $p/q$, showing the windings of the rational strip around the unit torus, $q$ times and $p$ along the $x$ and $y$ directions respectively. To fix $\phi_S$ we require a) that the total flux for the path shown in Fig.\ref{fig:cutproj}d), namely, $(p+q)\phi^S$ be an integer. This means that $\phi^S \propto (p+q)^{-1}$.  We require next b) that it lead to the hierarchical structure of Fibonacci chain spectra, as seen by explicit computation in a perturbative limit \cite{kkl,niunori,piechon}. In the Quantum Hall problem, we recall a result concerning sub-bands of the spectrum due to Hofstadter \cite{hofstadter}. It states that, for a rational flux $f$, the effective flux governing sub bands is $\phi'=[f^{-1}]$. Now, setting $f= F_{k-1}/F_{k}$, then using the recursion relation for Fibonacci numbers, it is easy to see that
\begin{eqnarray}
[f^{-1}] =  \frac{F_{k-2}}{F_{k-1} }
\end{eqnarray}
in other words, the effective flux has a value associated with the next smallest approximant. This choice of flux assures recursivity of the band structures as one goes between successive approximants. Henceforth, we take $\phi^S_k= F_{k-1}/F_{k}$ for the $k$th approximant. In the quasiperiodic limit it converges to the (inverse) golden mean, $\tau^{-1}$ \cite{foot2}.

Working in the Landau gauge with a vector potential $\vec{A}=(0,Bx,0)$, the 2D Quantum Hall Hamiltonian can be expressed in the real space basis as  
\begin{eqnarray}
    H^{2D} = \sum_{m,n}  t_a c^\dag_{m+1,n}c_{m,n} + h.c. \nonumber \\
    + \sum_{m,n} t_b(m) c^\dag_{m,n+1}c_{m,n} + h.c. 
    \label{eq:ham2D}
\end{eqnarray}
where $c^\dag_{mn} $ ($c_{mn}$) are the electron creation (destruction) operators on the lattice point $\vec{r}_{mn} = (m\vec{x}+n\vec{y})a$.  The vertical hopping amplitudes are position dependent $t_b(m) = t_b ~e^{2i\pi m \phi^S}$ (dropping the subscript $k$ in the flux for simplicity). Thanks to the chiral symmetry of the problem, one can consider $t_a$ and $t_b$ both to be positive, without loss of generality. To solve for the spectra, toroidal boundary conditions are imposed. The magnetic unit cell is, as in the standard QH problem, larger than the structural unit cell. This bigger cell is obtained by taking $N=F_k$ copies of the basic structural unit cell along each of the two directions defined by the primitive basis vectors as illustrated in Fig.1c). Solving this problem, one obtains the well-known QH energy spectrum consisting of $N$ energy bands separated by $N-1$ gaps (note that for even values of $N$, two of the bands meet at $E=0$).   

In order to extrapolate from 2D to 1D limit, the inter-chain couplings between different chains are multiplied by a factor $0\leq\epsilon\leq 1$. Fig.\ref{fig:cutproj}b) shows the square lattice, with the intrachain bonds in blue, and inter-chain bonds in red.
The resulting $\epsilon$-dependent ``Fibonacci-Hall" Hamiltonian is:
\begin{eqnarray}
\label{eq:hamHtoF}
    \langle i\vert H^{FH}\vert j\rangle &=& \langle i\vert H^{2D}\vert j\rangle \qquad i,j \in C   \nonumber \\
    \langle k\vert H^{FH}\vert l\rangle &=& \epsilon \langle k\vert H^{2D}\vert l\rangle \qquad k\in C, l\in C'\neq C  \nonumber \\
\end{eqnarray}
where the first relation holds for sites $i$ and $j$ that are located in the same chain (denoted by $C$), and the second line holds for sites located on two different chains. 


To resume, $H^{FH}$ depends on several parameters. The approximant size $F_k$, firstly, fixes the value of flux $\phi^S$. The explicit parameters of the Hamiltonian are $t_a$, $t_b$ and $\epsilon$ of which only two are independent, (we can set $t_a=1$ without loss of generality). We will not discuss the extension of the model to include onsite potentials,  but this could be done in an analogous way. There is an additional implicit parameter namely, the position of the selection strip. This parameter has no incidence on the band structure when periodic boundary conditions are applied but does play a role in the energies of the edge states in an open system. Nb. the $H^{FH}$ model for zero ``flux" is topologically trivial but expected to have electronic properties different from the square Fibonacci model \cite{lifshitz,thiem} (a compact expression for our Hamiltonian using the so-called conumbering scheme \cite{siremoss} is given in the S.I. \cite{supple} ).   

\section{Phase diagram of the Fibonacci-Hall model}
This section discusses the Fibonacci-Hall model in the $\epsilon - \delta$ parameter space, where $\delta=t_b-t_a$. The original 2D QH model is obtained for $\epsilon=1$. A set of identical decoupled Fibonacci Hamiltonians given by Eq.\ref{eq:hamFibo} is obtained for $\epsilon=0$. In this strict 1D limit, the magnetic flux dependence is removable by a gauge transformation (the physical reason is that there are of course no closed loops in this limit, thus magnetic flux plays no role). Both these limiting cases are well-understood.  In all the following we will consider approximants, that is, periodic systems. The quasiperiodic limit $k \rightarrow \infty$ is well-understood for these limiting models. When $\epsilon=0$, the 1D quasicrystalline model is critical for all $\delta\neq 0$, with multifractal spectrum and states (see the review in \cite{jagaRMP}). The 2D QH model is only critical for $\delta=0$ \cite{jitomirskaya}, the isotropic case. 

In the QH problem, a change occurs in the spectrum when one moves away from the isotropic limit $t_a=t_b$. Zhang et al \cite{bulmash}, considering the semi-classical limit, show that the Fermi surfaces open along one direction giving rise to what they called ``almost mobility edges" separating strongly localized from weakly localized states. Similar changes occur in the spectrum of $H^{FH}$ when $\epsilon$ is reduced for $\delta$ fixed, except that the symmetry is changed from square to rhombic (along new tilted axes).  Numerical results for the evolution of energy levels of $H^{FH}$ for the $k=4$ approximant are shown in Fig.\ref{fig:bands}a). For clarity, we have shown the energy eigenvalues solved only at four special points $(0,0), (\pi,0),(0,\pi)$ and $(\pi,\pi)$ in reciprocal space. These correspond to extremal states of the three bands, and are the energies for which band crossings occur as $\epsilon$ is varied. In our multigap system, states cross at different values of $\epsilon$ for each gap. The main gap crossing occurs for the smallest value, denoted by $\epsilon_c$. 


The value of $\epsilon_c$ as a function of $\delta$ can be approximately computed using a 3-band model. Indeed, the recursive band structure of the Fibonacci system implies that the parent chain for the two main gaps is the 3-site approximant chain $t_a-t_a-t_b$. 
For zero flux, the Hamiltonian in the reciprocal space basis of this tilted system is given by a $3 \times 3$ matrix , 
\begin{eqnarray}
  \begin{pmatrix}
       0 & 1 + \epsilon \rho e^{iK'} &  e^{-iK} + \epsilon \rho e^{i(-K+K')} \\
    1 + \epsilon \rho e^{-iK'} &  0 & \rho + \epsilon  e^{-iK'}  \\
     e^{iK} + \epsilon \rho e^{i(K-K')} & \rho + \epsilon  e^{iK'} & 0 \\ 
   \end{pmatrix} \nonumber
\end{eqnarray}
where $\rho=t_b/t_a=1+\delta$ and the reduced momentum components $K$ and $K'$ vary within $[0,\pi]$. Diagonalization yields the three energy bands in momentum space, $E_j(K,K',\epsilon)$ ($j=1,2,3$), and one can now study band crossings. The lower band gap is traversed by two branches, $E_2(0,\pi,\epsilon)$ and $E_1(0,0,\epsilon)$, that intersect for $\epsilon=\epsilon_c$. The energy surfaces also show a change of form from compact to extended, at this point marking a Lifshitz transition.  Series expansion for small $\epsilon$ and $\delta$ gives the solution $\epsilon_c \approx 2\delta/(3+2\delta)$. For larger $\delta$ one can solve for $\epsilon_c$ numerically. A study of the $k=3$ and also bigger approximants shows similar behavior. For any system size, the following conjecture
\begin{eqnarray}
\epsilon_c= \frac{\delta}{(\phi^S)^{-1}+\delta }  
\label{eq:crit}
\end{eqnarray}
is well obeyed by the data (see S.I. \cite{supple}). 
This Lifshitz transition is indicated in Fig.\ref{fig:bands}b) by the dashed blue lines demarcating the topological and trivial regions. 

Level crossings in the other gaps could, in principle, be similarly described via a theory utilizing suitably renormalized hopping parameters (discussed in the strong modulation limit in \cite{kkl,niunori,piechon}). A complete study of all these transitions is outside the scope of this paper but we find that, just as for the main gap, no level crossings occur when $\delta=0$. One can thus reach the off-diagonal Fibonacci chain model starting from the QH problem without any band crossings, in a finite system as follows:

\begin{figure*}
\includegraphics[width=0.25\textwidth]{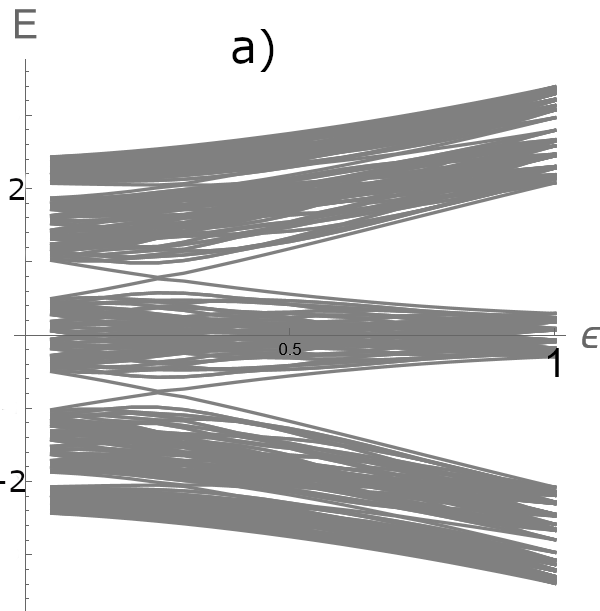}
\includegraphics[width=0.2\textwidth]{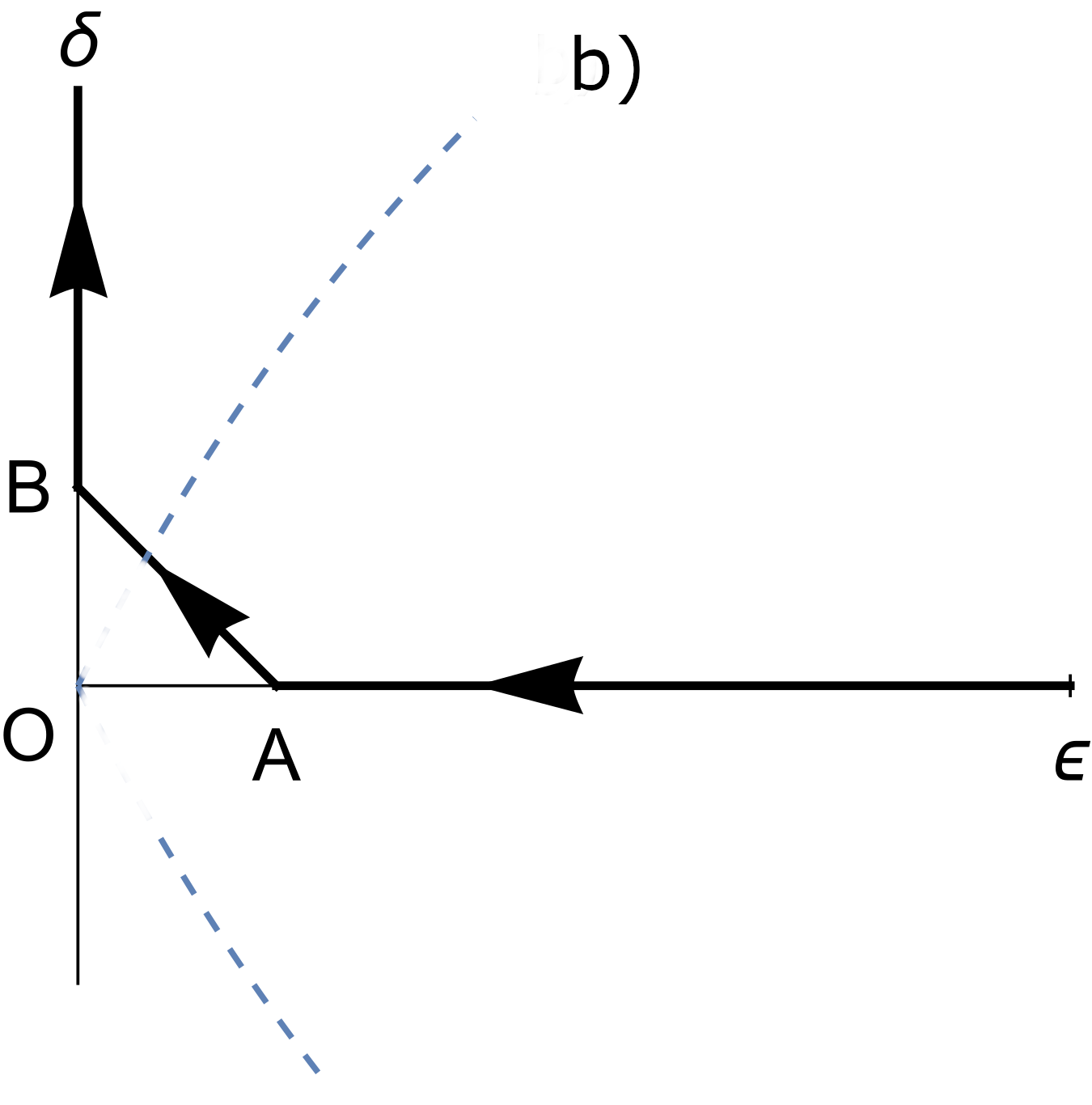}
\includegraphics[width=0.5\textwidth]{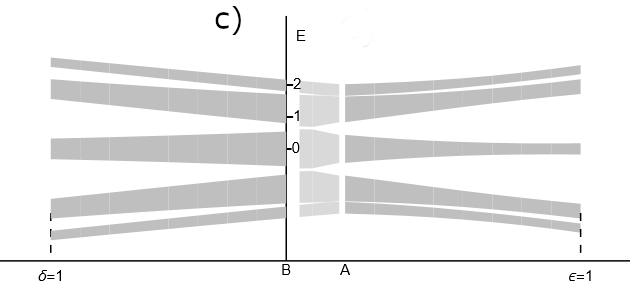}
\caption{a) The evolution of energy levels of $H^{FH}$ as a function of $\epsilon$, showing crossings for fixed $\delta=0.5$. Data was obtained for $k=4$ approximant (125 site magnetic unit cell) for four special points in reciprocal space $(0,0), (\pi,0),(0,\pi)$ and $(\pi,\pi)$.  b) Schema of the path in $\delta$-$\epsilon$ parameter space used for the plot in figure c). The point A corresponds to a relatively large value of $\epsilon_{min}=0.2$ so that the gaps can be seen more clearly. The dashed blue lines indicate the Lifshitz transition for the main gap Eq.8.  b) Evolution of bands and gaps of the Hamiltonian Eq.\ref{eq:hamHtoF} for a $N=5$ approximant when going adiabatically from the isotropic QH model ($t_b=\epsilon=1$) to the off-diagonal Fibonacci model with $\epsilon=0,t_b=2$.  }
\label{fig:bands}
\end{figure*}

\begin{enumerate}
\item[--] Keeping $\delta=0$, $\epsilon$ is reduced from 1 to a small value $\epsilon_{min} \neq 0$ of the order of the typical energy level spacing, which decreases with system size. 
\item[--] When the point A at $(\epsilon_{min},0)$ is reached, one goes from A to B at $(0,\delta_{min})$ (with $ \delta_{min} \sim \epsilon_{min} $), thus avoiding the origin. 
\item[--] From B, one goes along the vertical axis to the desired final value of $\delta$. 
\end{enumerate}
Fig.\ref{fig:bands}b) shows the path in parameter space and the evolution of bands along the path is shown in Fig.\ref{fig:bands}c) for the $k=4$ approximant. It can be seen that the gaps remain open all through. For purposes of clarity, in order to see gaps clearly, we have chosen a large value of $\epsilon_{min}=0.2$.

The Chern numbers for each of the bands, which here are the same as the Bott indices, can be conveniently computed using the real space method due to Loring and Hastings \cite{loring}. Site positions $\vec{r}_i$ are re-expressed as a pair of variables $(\theta_i,\varphi_i)$ in the range $[0,2\pi)$, used in defining Hermitian matrices $\Theta = \exp(i \theta)$ and $\Phi = \exp(i \varphi)$. Using the projection operators into the $n$th band $P_n$, one computes the matrices $U = P_n\Theta P_n$ and $V = P_n\Phi P_n$. For a large enough system size these are almost unitary because of the short range nature of the Hamiltonian \cite{loring}. The Chern number of the $n$th band is given by
\begin{eqnarray}
\mathcal{C}_n = \frac{1}{2\pi} \mathrm{Im} [\mathrm{Tr Log} (VUV^\dag U^\dag)]
\end{eqnarray} 
The gap labels are given by
\begin{eqnarray}
    g_n= \sum_{m=1}^n\mathcal{C}_m
    \label{eq:gapchern}
\end{eqnarray}
Table 1 lists the Chern numbers and gap labels found using Eq.\ref{eq:gapchern} for the $N=8$ approximant. In an approximant gaps are of two kinds: stable gaps, having a fixed label as $k\rightarrow \infty$, or unstable (varying gap label) \cite{macegaplabels}. For the approximant shown in Table 1, the central $g=4$ gap is the only unstable gap. Note that the computed values of $\mathcal{C}_n$ are exact. These values remain constant when $\epsilon$ is reduced from 1 to a minimum value of $\sim 0.15$, when they drop to 0. The gap labels we obtain in this way agree with the values determined using the gap labeling theorem Eq.\ref{eq:gaplabeling}.  

\vskip 1cm
\begin{table}
\begin{tabular}{|c|c c c c c c c c |}
\hline 
$n$ & 1 & 2 & 3 & 4 & 5 & 6 & 7 & 8 \\
\hline 
&&&&&&&&\\
$\mathcal{C}_n$ & -3 & 5 & -3 & 5 & -3 & -3 & 5 & -3 \\
&&&&&&&&\\
\hline
&&&&&&&&\\
$g_n$     & -3 & 2 & -1 & 4 & 1  & -2 & 3&    \\ 
&&&&&&&&\\
\hline
\end{tabular}
\caption{Upper row: the Chern numbers $\mathcal{C}_n$ of all the bands of $H^{FH}$ (Eq.\ref{eq:hamHtoF}) for system size $N=8$. The integer values are exact. They are constant as $\epsilon$ varies, down to a minimum non-nul value. Lower row : the gap labels for each of the 7 gaps deduced from the upper row using Eq.\ref{eq:gapchern}. The label $g=4$ corresponds to an unstable gap, all the other gaps are stable \cite{macegaplabels}. }
\end{table}

\section{Edge states for $\delta\neq 0$ and Lifshitz transition}

In open systems the bulk-edge correspondence leads to edge modes, which can be used as markers for the various topological transitions in the $\epsilon-\delta$ plane. We considered square unit cells (as discussed in the S.I. \cite{supple}) and solved for the eigenstates of Eq.\ref{eq:hamHtoF} when the long edges are left open, and periodic boundary conditions are applied along the $y'$ direction. Here we study edge states in the main gap for fixed finite $\delta$. When $\epsilon=1$, one is in the topological sector, with an edge state.  As $\epsilon$ is reduced, a topological transition occurs, with disappearance of the edge state and simultaneously a change of the Chern number. 

\begin{figure*}[h!]
\includegraphics[width=0.3\textwidth]{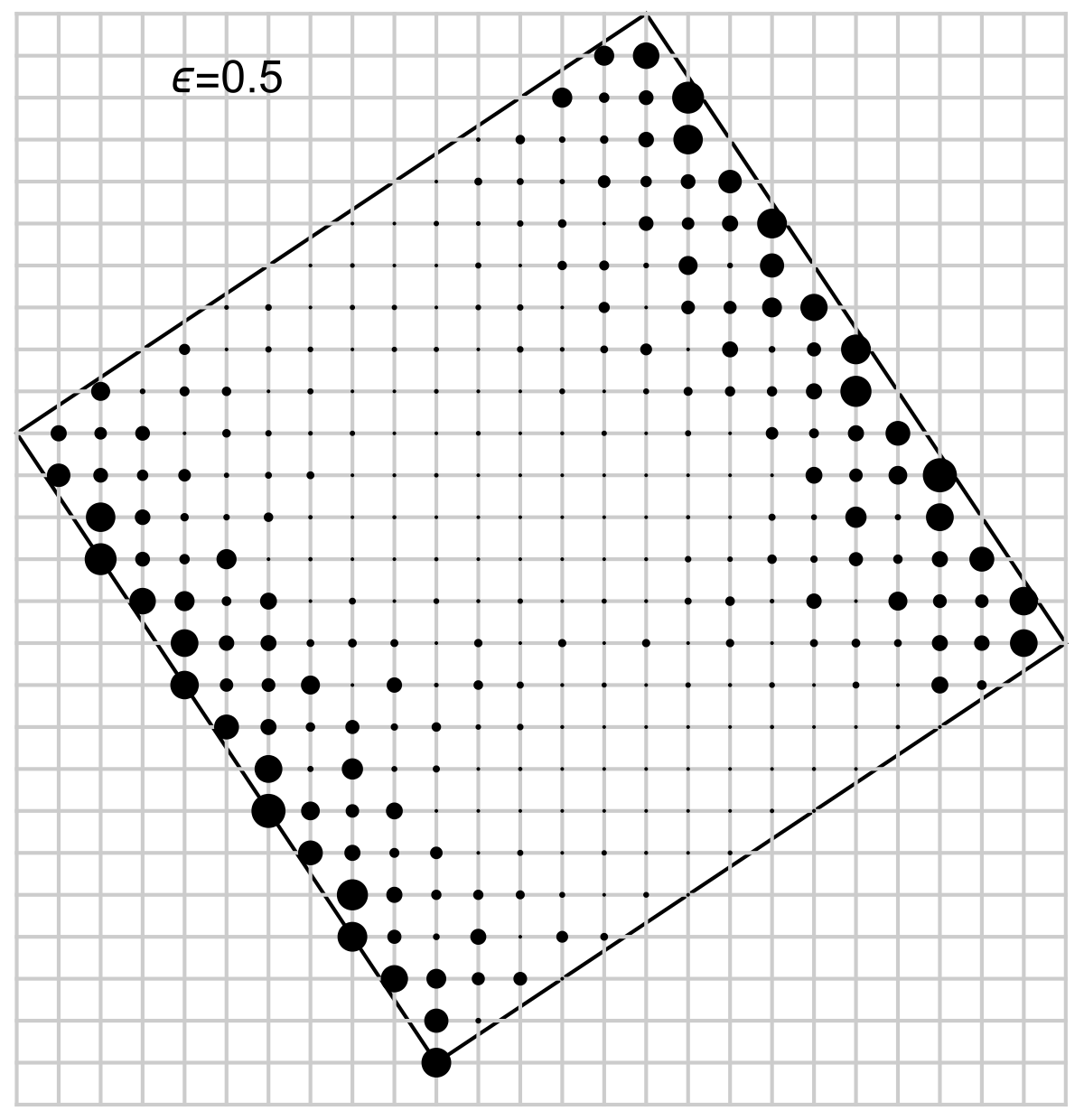} 
\includegraphics[width=0.3\textwidth]{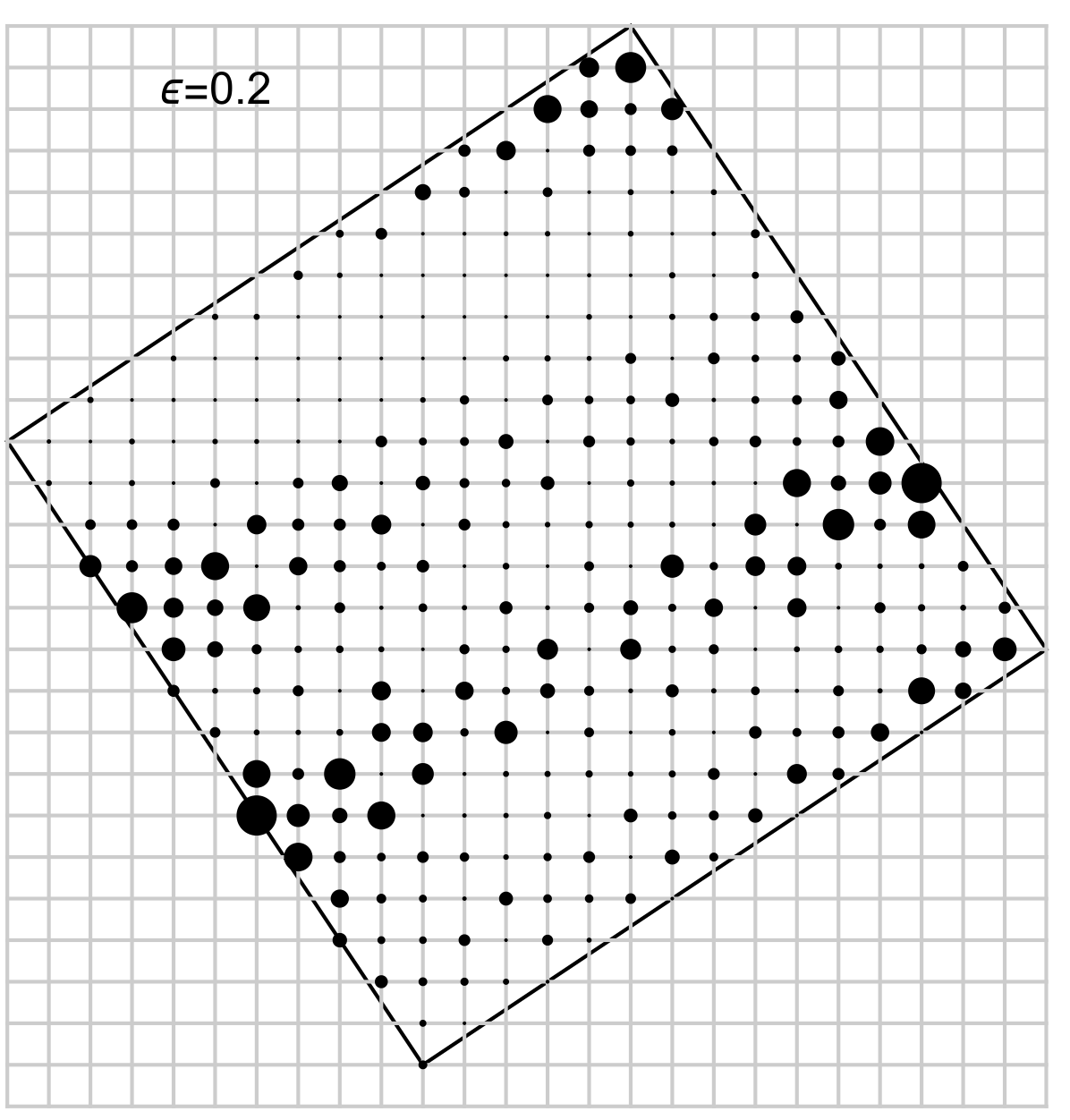}
\includegraphics[width=0.3\textwidth]{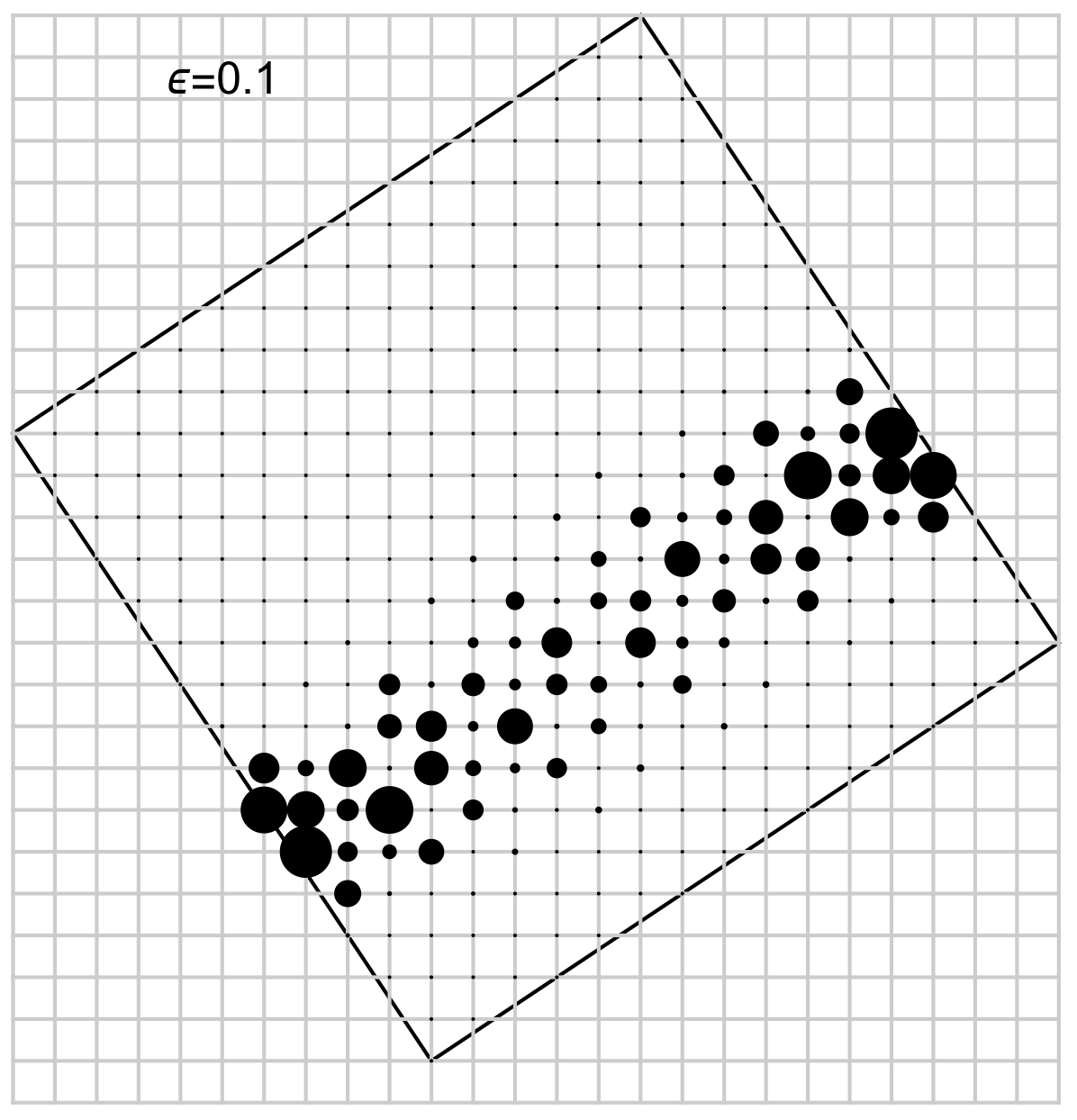} 
\caption{ Edge state within the $g=-1$ gap, for $\epsilon$ values above (left), at (center) and below (right) the transition at $\epsilon \approx 2$. A square unit cell for the $N=5$ approximant was used with periodic boundary conditions along the $y'$ direction ($\delta=0.4$). The radii of circles are determined by the amplitudes of the wavefunction on the sites.}
\label{fig:edgestates}
\end{figure*}

This evolution of edge states is shown in Figs.\ref{fig:edgestates}. The computations are done for $\delta=0.4$ and a fixed band filling. The spatial profile of the state lying in the one of the big gaps (label $g=-1$) is shown using circles of size proportional to the onsite wavefunction amplitude. The state is shown for three values of $\epsilon$ above, at and below the topological transition. The total Chern number of the filled bands stays equal to 1 as $\epsilon$ is reduced, then jumps discontinuously to 0 at the transition at $\epsilon\approx 0.2$, in agreement with the formula of Eq.\ref{eq:crit}. Below this value, the edge state is no longer localized on the open 2D edge. It is localized around a single chain, and resembles a bulk 1D state.  
The principle of bulk-edge correspondence thus holds: the 2D edge state that exist in the topological sector disappears in the trivial sector ($\epsilon < \epsilon_c$). Another discontinuous change occurs when $\epsilon=0$. In this limit, 1D edge states would appear, as expected theoretically and confirmed by experiments \cite{tanese}. 

\section{Conclusions}
This paper considers the relation between the 2D Quantum Hall problem and 1D Fibonacci quasicrystals that was pointed out in a number of pioneering studies \cite{kraus}, and proposes a ``missing link" between them, the Fibonacci-Hall model. The Chern numbers are seen to result from  a ``geometric flux" parameter $\phi^S$ arising due to the winding of the quasicrystal in 2D space. This connection clarifies how the 1D quasicrystal can ``inherit" topological properties from the 2D system.  After describing its construction, and outlining its phase diagram, we computed its Chern numbers. From them we deduce gap labels for the Fibonacci chain, in accord with those determined by the gap labeling theorem, and as verified by experimental investigations \cite{tanese}. The topological invariants are robust with respect to perturbations providing that gaps do not close. In disordered Fibonacci approximants, it is known that the gaps structure  -- and therefore the gap labels -- are preserved for sufficiently weak disorder \cite{tarzia}. 
In an alternative scenario, Juricic et al \cite{juricic} have related topological invariants of the 1D system to those of the parent 2D Chern insulator, by computing local Chern numbers in an open system. The approach views quasicrystals as holographic images of higher-dimension topological crystals on lower-dimensional branes. 

Similar models can be written for other 1D quasicrystals, for example the silver and other metallic mean chains \cite{footnote}. It will be even more interesting to consider the 4D Quantum Hall problems related to 2D quasicrystals such as the Ammann-Beenker tiling \cite{grunbaum,AJduneau}. The spectral gaps for this tiling have been discussed in \cite{jaga2023}, along with a labeling scheme.  New physical properties can be studied for the higher dimensional topological system, such as non-linear responses involving first and second Chern numbers. It would be an extension to a real 2D case of the pioneering work of Lohse et al \cite{lohse} on quantized nonlinear transport in a synthetic 2D quasicrystal obtained by multiplying two Aubry-André chains.   

{\section{Acknowledgments}} I gratefully acknowledge many useful discussions with Pavel Kalugin, Frédéric Piéchon, Gilles Montambaux and Mark Goerbig.



\begin{thebibliography}{15}
\bibitem{foot2} Due to the symmetry of the problem under $\phi \rightarrow 1 - \phi$, an equivalent choice for the flux would be $F_{k-2}/F_k$.
\bibitem{footnote} The golden, silver and other metallic means are irrational numbers that are solutions of the quadratic equation $x^2-nx-1=0$, for $n=1,2,..$
\end{thebibliography}
\end{document}


\title{Supplementary Information for Missing link between the 2D Quantum Hall problem and 1D quasicrystals}

\author{Anuradha Jagannathan}
\affiliation{
Laboratoire de Physique des Solides, Universit\'{e} Paris-Saclay, 91405 Orsay, France
}
\date{February 2023}
\maketitle

\section{Fibonacci-Harper model for zero flux}
This section describes the Fibonacci-Harper model in zero magnetic field, $H^{FH}_{\phi^S=0}$. The Hamiltonian of Eqs. 5,6 and 7 of the main text can be written as the sum of two nearest neighbor Hamiltonians, $H^{FH}_{\phi^S=0} = H_{\|} + \epsilon H_{\perp}$. Here $H_{\|}$ groups together the hopping terms along a given chain, and $H_{\perp}$ the hopping terms between chains. A notable feature of this model is, that the roles of $t_a$ and $t_b$ are interchanged when going from the parallel to the perpendicular direction. A convenient basis to represent the Hamiltonian consists of using the so-called conumber basis, which numbers the sites according to increasing $y'$ coordinates. One can write $H^{FH}_{\phi=0}=H_{\|} + \epsilon H_{\perp}$ as already described in the main text. $H_{\|}$ is nothing but the Hamiltonian of a periodic approximant, which is known to have a T\"oplitz form \cite{siremoss}. With periodic boundary conditions along the chain, it is the  $F_k \times F_k$ matrix 

\begin{equation}
 H_{\|}= \begin{pmatrix}
    0 & & t_ae^{iK} &..& t_b & ..\cr
    & 0 & & t_a & .. & t_b & ..\cr
    && 0& & t_a& .. & t_b & ..\cr
    &&&&&&&&\ddots\cr
    &&&&&&&&\ddots\cr
    &&&&&&\ddots\cr
    t_ae^{-iK}&&&&&\cr
     & t_a&&&&&\cr
    &  & t_a&\cr
    t_b & & & \cr
    & t_b&& & \cr
    &&&&\ddots\cr
     &&&\ddots\cr
  \end{pmatrix}
\end{equation}

where the only non-zero matrix elements, $t_a$ and $t_b$, lie on diagonals located at distances $F_{k-2}$  and $F_{k-1}$ from the main diagonal. The reduced momentum variable $K$ lies in the interval $[0,\pi]$. Similarly, $H_{\perp}$ can be written as

\begin{equation}
 H_{\perp}= \begin{pmatrix}
    0 & & \tilde{t}_be^{iK} &..& \tilde{t}_a  & ..\cr
    & 0 & & \tilde{t}_b  & .. & \tilde{t}_a & ..\cr
    && 0& & \tilde{t}_b & .. & \tilde{t}_a  & ..\cr
    &&&&&&&&\ddots\cr
    &&&&&&&&\ddots\cr
    &&&&&&\ddots\cr
    \tilde{t}_be^{-iK}&&&&&\cr
     & \tilde{t}_b &&&&&\cr
    &  & \tilde{t}_b &\cr
    \tilde{t}_a^* & & & \cr
    & \tilde{t}_a^* && & \cr
    &&&&\ddots\cr
     &&&\ddots\cr
  \end{pmatrix}
\end{equation}

with $\tilde{t}_a = t_a e^{iK'}$ where the momentum variable $K'$ lies in the interval $[0,\pi]$. 

As a result, the total Hamiltonian also has a rather simple T\"oplitz form. In particular, for $K=K'=0$ one has 

\begin{equation}
 \begin{pmatrix}
    0 & & t_1 &..& t_2 & ..\cr
    &0 & & t_1 & .. & t_2 & ..\cr
    &&0& & t_1& .. & t_2 & ..\cr
    &&&&&&&&\ddots\cr
    &&&&&&\ddots\cr
    t_1&&&&&\cr
    . & t_1&&&&&\cr
    .& . & t_1&\cr
    t_2 & & & \cr
    & t_2&& & \cr
    &&&&\ddots\cr
     &&&\ddots\cr
  \end{pmatrix}
\end{equation}
where $t_1=t_a+\epsilon t_b$ and $t_2=t_b+\epsilon t_a$. 

The eigenvalues for this matrix are those of a 1D approximant chain with hopping parameters $t_1$ and $t_2$. This leads to the conclusion that, at the origin $(K,K')=(0,0)$, this spectrum will be multifractal for $t_1\neq t_2$ in the $k\rightarrow \infty$ limit. The study of the full spectrum is left for future work. 

Finally, note that for $\epsilon=1$, this problem becomes translationally invariant and one has the standard dispersion relation $E(k_x,k_y) = 2t_a \cos(k_x)+ 2t_b \cos(k_y)$. 

\section{Band structure of $H^{FH}$ for $\delta\neq 0$ }
This section shows the behavior of individual levels of the Hamiltonian $H^{FH}$ when the parameter $\epsilon$ is varied, in the anistropic case $t_a \neq t_b$.  Periodic boundary conditions were imposed and the diagonalization carried out for four extremal points in $\vec{K}$ space -- $(0,0), (\pi,0), (0,\pi)$ and $(\pi,\pi)$ in dimensionless units. 
 This transition represents a variant of the Lifshitz transition of the type discussed in \cite{bulmash}. At such crossings, states become delocalized along one direction, namely parallel to the chains. 
We have seen in the main text using a series expansion that the main gap crossing should occur approximately for $\epsilon_c = \vert 2\delta/(3+2\delta)\vert = \delta/(\frac{3}{2}+\delta)$. This leads to a conjecture -- that for larger approximants the transition should occur for
   \begin{eqnarray}
 \epsilon_c = \vert \frac{\delta}{\tau + \delta} \vert   
 \label{eq:crit2}
\end{eqnarray}
where $\tau = \lim_{k\rightarrow \infty} F_k/F_{k-1}$ is the golden mean. 

\begin{figure*}[h]
\includegraphics[width=0.5\textwidth]{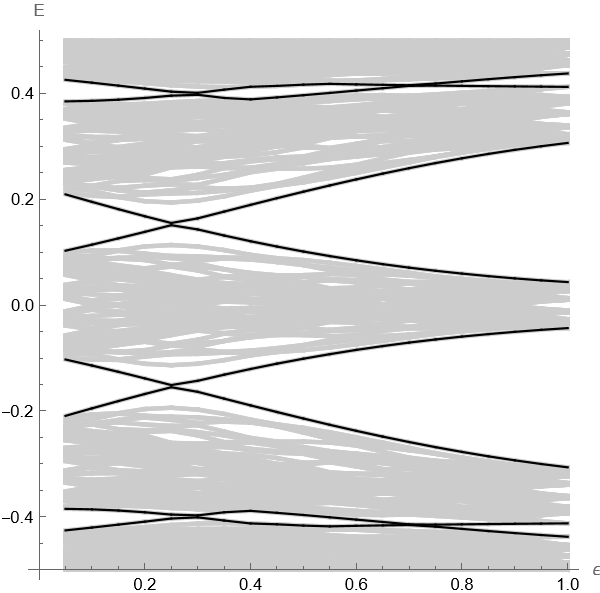}\hskip 2cm
\includegraphics[width=0.5\textwidth]{SIfig1b.png}
\caption{  top) The evolution of energy levels of $H^{FH}$ for fixed $\delta=0.5$ as a function of $\epsilon$, showing crossings. Data was obtained for $k=4$ approximant (125 site magnetic unit cell), for four points $(0,0), (\pi,0),(0,\pi)$ and $(\pi,\pi)$ in reciprocal space. Level-crossings across the four band gaps are shown outlined in black. The band width has been normalized to unity. bottom) Plot of the critical values $\epsilon_c$ versus the renormalized hopping amplitude $t_b$. Blue points are data for $k=4$, orange points are data for $k=5$. The orange dashed line is the theoretical expression obtained by series expansion of $k=3$ model. The grey dashed line is the conjecture of Eq.\ref{eq:crit2}.}
 
\label{fig:crossings}
\end{figure*}

Fig.\ref{fig:crossings}a) shows results obtained for the $k=4$ approximant. This is a system having 125 sites in its magnetic unit cell. The parameter $\delta=0.5$. Data for the critical value $\epsilon=\epsilon_c$ in the case of the main gap is plotted in Fig.\ref{fig:crossings}b) for two different approximants ($k=4,5$). Level-crossings across the four band gaps are shown outlined in black. Fig.\ref{fig:crossings}b) is a plot of $\epsilon_c$ for the largest bandgap versus the renormalized hopping amplitude $t_b$ for different system sizes. One observes that the data are well fitted by the conjectured expression Eq.\ref{eq:crit2}, which is shown by the grey dashed line in the figure. The orange dashed line is the expression obtained by the power law expansion for $k=3$ system that we discussed in the main text.

\section{Rectangular unit cells for approximant structures.}
To study edge states we have considered rectangular unit cells whose edges lie parallel to the $x'$ and $y'$ axes. The periodicity of the system along the $x'$ axis is unchanged -- the structure repeats after $F_k$ sites. However the periodicity along $y'$ is larger by a factor $F_k$ times. If we consider an approximant of slope $F_{k-2}/F_{k-1}$ the two lattice vectors are $\vec{A}=(F_{n-1},F_{n-2})$ (oriented along the $x'$ axis) and $\vec{B}= F_k(-F_{k-2},F_{k-1})$ (along the $y'$ axis). The magnetic unit cell is obtained by taking 5 copies of this cell along the $x'$ direction. This is the square cell shown in Fig.3a) of the main text.

\begin{figure*}
\includegraphics[width=0.3\textwidth]{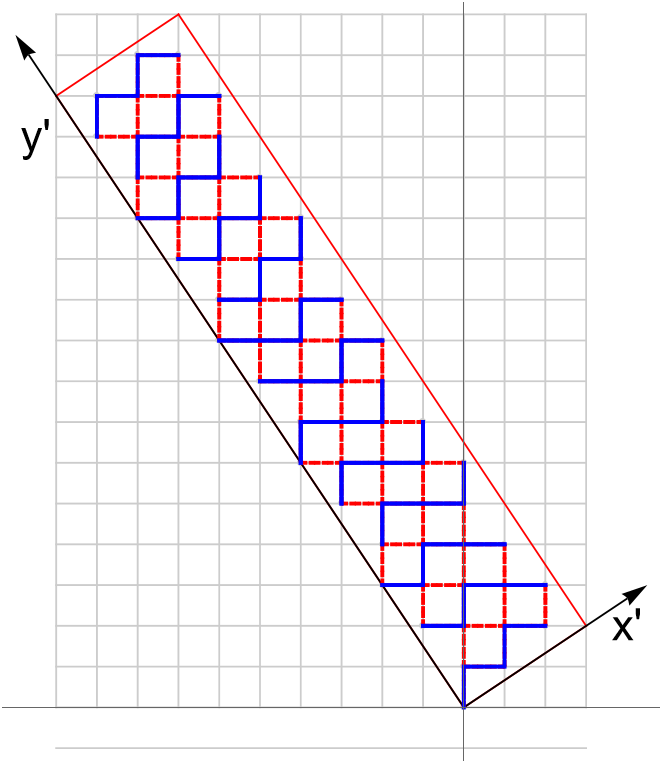}
\caption{A unit cell with rectangular form for $k=4$ approximant, having 13 parallel chains of 5 sites. Blue (red) bonds show intra (inter)chain hoppings between sites. The magnetic unit cell is obtained by taking 5 copies of this cell along the $x'$ direction. }
\label{fig:rectcell}
\end{figure*}

An example of a rectangular cell is shown in Fig.\ref{fig:rectcell} for the approximant $k=4$ of 5 sites. The unit cell contains 13 parallel Fibonacci approximant chains -- shown in blue -- coupled to each other (when $\epsilon$ is finite). As can be seen from the figure, when the boundaries along the long edges are opened, these finite chains can differ from each other by an origin shift. Thus, in the 1D limit, edge states appear in the gaps at different energies for different chains of the open system.

